\documentclass[12pt,preprint]{emulateapj}

\usepackage{graphicx}
\usepackage{amsmath}

\begin{document}

\title{The cosmic distance duality relation with strong lensing and gravitational waves: an opacity-free test }
\author{Kai Liao$^{1}$}
\affil{
$^1$ {School of Science, Wuhan University of Technology, Wuhan 430070, China.}}
\email{liaokai@whut.edu.cn}

\begin{abstract}
  The cosmic distance duality relation (CDDR) is a fundamental rule in cosmological studies.
  Given the redshift $z$, it relates luminosity distance $D^L$ with angular diameter distance $D^A$ through
  $(1+z)^2D^A/D^L\equiv1$. Many efforts have been devoted to test CDDR with various observational approaches.
  However, to the best of our knowledge, those methods are always affected by cosmic opacity which could make CDDR violated due to the non-conservation of photon number.
  Such mechanism is more related with astropartical physics.
  In this work, to directly study the nature of space-time, i.e., to disentangle it from astropartical physics, we propose a new strategy to test CDDR with strong lensing providing $D^A$
  and gravitational wave providing $D^L$. It is known that the propagation of gravitational wave is unaffected by cosmic opacity.
  We demonstrate distances from optical lensing observation are also opacity-free.
  These two kinds of distance measurements make it possible to test space-time.
  Our results show the constraints on the deviations of CDDR will be very competitive to current techniques.

\end{abstract}
\keywords{cosmology: distance scale - gravitational lensing: strong - gravitational waves }

\section{Introduction}
The development of modern cosmology strongly relies on the measured distance-redshift relation.
While the redshift ($z$) of a celestial object is relatively easy to accurately get with spectral lines,
cosmological distance measurements are significantly important. In a general space-time, the direct observational quantities are
luminosity distance $D^L$ and angular diameter distance $D^A$. Theoretically, if the following three conditions are satisfied:
\begin{itemize}
\item 1. The space-time is described by a metric theory of gravity;
\item 2. Photons travel along null geodesics;
\item 3. Photon number is conserved,
\end{itemize}
then $(1+z)^2D^A/D^L\equiv1$ called the ``cosmic Distance Duality Relation" (CDDR) holds~\citep{Etherington1933}. Note that Condition 1 and 2 are related with
the nature of space-time and more fundamental, while Condition 3 usually corresponds to astrophysical mechanisms or particle physics.
Testing the validation of CDDR with observation would
either strengthen our current knowledge on the Universe or reveal new physics/astrophyiscal mechanisms~\citep{Bassett2004}.
Various methods have been used to test CDDR.

To perform a test on CDDR, one needs a luminosity distance measurement plus an angular diameter distance measurement at the same redshift.
For example, the most commonly used combination consists of $D^L$ data from type Ia supernovae (SNe Ia) as the standard candles and
$D^A$ data from the galaxy clusters~\citep{Bernardis2006,Holanda2010,Li2011,Holanda2012,Hu2018}. It has been conjectured that the cosmological dust might make the observed SNe Ia dimming~\citep{Lima2011}.
Other dimming mechanisms include extragalactic magnetic fields turning photons into light axions, gravitons, Kaluza-Klein modes associated with
extra-dimensions or a chameleon field. They are all taken as the cosmic opacity~\citep{Avgoustidis2010,Liao2015a} which could change the $D^L$ measurements
leading to violation of CDDR.
Meanwhile, $D^A$ from galaxy clusters are based on the SZE+X-ray surface brightness observations~\citep{Uzan2004}:
\begin{equation}
D^A\propto \frac{\Delta T^2_{CMB}}{S_X},
\end{equation}
where $\Delta T_{CMB}$ is the temperature change when the cosmic microwave background (CMB) radiation
passes through the hot intra-cluster medium (Sunyaev-Zel'dovich effect). $S_X$ is the X-ray surface
brightness of galaxy cluster. They are both affected by the cosmic opacity since the measurements are intensity quantities~\citep{Li2013}.
In other methods, $D^L$ can come from Gamma Ray Bursts (GRBs) at high redshifts~\citep{Holanda2017}. Likewise, luminosity distances of GRBs also depend on cosmic opacity.
$D^A$ can come from ultra-compact radio sources~\citep{Li2018} and baryon acoustic oscillations (BAOs)~\citep{Wu2015}, however, they either suffer from cosmic opacity or
assume a $\Lambda$CDM making the test model-dependent.

If one wants to exclude the impacts by the cosmic opacity (also the cosmological models), i.e., to
model-independently test the space-time only, direct opacity-free distance
measurements should be applied.
On one hand, gravitational waves (GWs) as standard sirens were proposed to give the $D^L$~\citep{Yang2019,Qi2019,Fu2019}.
There are two benefits: firstly, the propagation of GWs is unaffected by cosmic opacity; secondly, they provide the direct
luminosity distances while SNe Ia in principle provide the relative distances.
On the other hand, strong lensing by elliptical galaxies were used to study cosmology~\citep{Chae2003,Oguri2007,Paraficz2009,Paraficz2010,Oguri2012,Cao2012,Tu2019,Chen2019,Wong2019}.
The angular diameter distance ratios from strong lensing observation carry the information of $D^A$~\citep{Liao2016,Yang2019}.
For an ideal model assuming the elliptical lens galaxy is described by a Singular Isothermal Sphere (SIS), once the Einstein radius ($R_E$)
and the central velocity dispersion $\sigma_v$ are measured by the separation angle of AGN multi-images and the spectroscopy, one can
infer the ratio of two angular diameter distances $D_{ls}^A/D_s^A\propto\theta_E/\sigma_v^2$, where the subscripts $l, s$ denote for lens and source, respectively.
However, the realistic lenses plus their environments are more complex~\citep{Jiang2007}. A universal simple model like SIS or its extensions for all lenses
can bring severe systematics~\citep{Xia2017}. More observational quantities and detailed analysis are required to model individual lensing systems one by one~\citep{H0LiCOW}.
Besides, when inferring $D^A$, a flat Universe has to be assumed~\citep{Liao2016,Holanda2016,Yang2019}. Furthermore, for a CDDR test, one needs two $D^L$ data at the same redshits of
the lens and source.

Current state-of-the-art lensing programs (for example, the H0LiCOW~\citep{H0LiCOW}) are focusing on time delay lenses.
With the measurements of time delays between AGN images, the central velocity dispersion of the lens, the host galaxy arcs plus lens galaxy imaging, and the mass fluctuation
along the line of sight (LOS), a good algorithm with blind analysis to control the systematics
can provide the ``time delay distances" which is a combination of three angular diameter distances $D_{\Delta t}=(1+z_l)D_l^AD_s^A/D_{ls}^A$
primarily depending on the Hubble constant~\citep{H0LiCOW}. Furthermore, time delay lenses were recently found to be more powerful for cosmological studies
with capability to measure the angular diameter distances to the lenses $D_l^A$~\citep{Paraficz2009,Jee2015,Jee2016,Yildirim2019,Wong2019}. The angular diameter distances
can be used in CDDR test~\citep{Rana2017}, though still in an opacity-dependent way with SNe Ia, whereas we will focus on disentangling the space-time nature from cosmic opacity with GWs.

In this work, we show that the angular diameter distance measurements from strong lensing are unaffected by
the cosmic opacity. Combining with the standard sirens, they could provide a direct opacity-free test on CDDR. Different from previous CDDR tests with lensing where
only $R_E$, $\sigma_v$ and $\Delta t$ are measured~\citep{Liao2016,Rana2017}, state-of-the-art and planned projects could measure individual lenses much better with
high-quality data from multiple aspects.

This paper is organised as follows: Section 2 introduces the $D^A$ from strong lensing and we give a explanation why it is opacity-free. Section 3 introduces
the $D^L$ from GW. We give the analysis and results in Section 4 and make a conclusion in Section 5. The flat $\Lambda$CDM with $\Omega_M=0.3$ and $H_0=70km/s/Mpc$
is assumed for simulating the lensing data and GW data.

\section{Angular diameter distances from strong lensing}
Strong lensing by elliptical galaxies has become a powerful tool to study astrophysics and cosmology~\citep{Treu2010}.
Systems with time delay measurements can yield a direct measurement of the angular diameter distance to the lens $D_l^A$~\citep{Jee2015,Jee2016}.

\begin{figure}
 \includegraphics[width=9cm,angle=0]{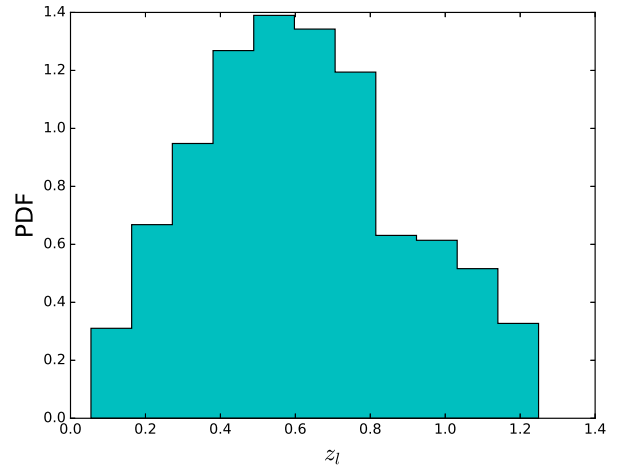}
  \caption{The selected redshift distribution of the lenses which can provide angular diameter distance measurements with $5\%$ precision.
  55 lenses are expected to be observed from current and upcoming projects.
  }\label{zl}
\end{figure}

\begin{figure}
 \includegraphics[width=9cm,angle=0]{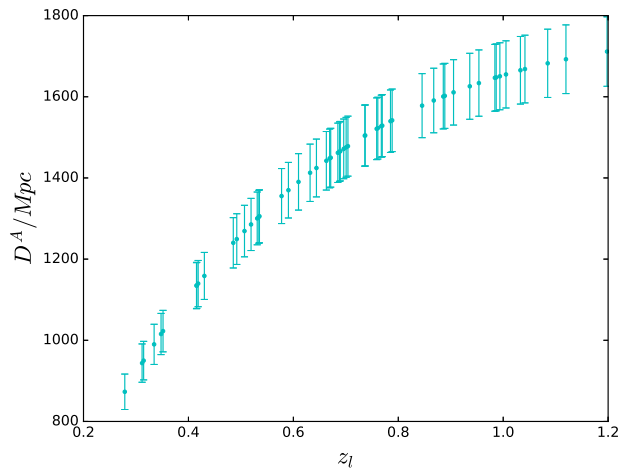}
  \caption{The mock $D^A$ data with uncertainty level $5\%$ from strong lensing.
  }\label{sl}
\end{figure}

Note that with very simple assumptions, the $D^A$ can be easily got~\citep{Rana2017}. However, we will introduce $D^A$ determination
on the standard of the H0LiCOW program~\citep{H0LiCOW}.
For illustration purpose, we briefly show how to measure the $D_l^A$ with observations.
The time-delay distance is given by:
\begin{equation}
(1+z_l)\frac{D_l^AD_s^A}{D_{ls}^A}=\frac{c\Delta t}{\Delta\phi(\boldsymbol{\xi}_{lens})},
\end{equation}
where $c$ is the speed of light. $\Delta t$ is the time delay measured by the light curves. $\Delta\phi=[(\boldsymbol{\theta}_i-\boldsymbol{\beta})^2/2-\psi(\boldsymbol{\theta}_i)-(\boldsymbol{\theta}_j-\boldsymbol{\beta})^2/2+\psi(\boldsymbol{\theta}_j)]$
is the Fermat potential difference for image angular positions $\boldsymbol{\theta}_i$ and $\boldsymbol{\theta}_j$, $\boldsymbol{\beta}$ denotes the source position, and $\psi$ is the two-dimensional lensing potential determined by the surface mass density of the lens
$\kappa$ in units of critical density $\Sigma_{\mathrm{crit}}=c^2D_{s}^A/(4\pi GD_{l}^AD_{ls}^A)$ through the Poisson equation $\nabla^2\psi=2\kappa$.
$\Delta \phi$ is determined by the lens model parameters $\boldsymbol{\xi}_{lens}$ which can be inferred with the high resolution imaging data.
Note that the LOS mass structure would also affect the measured time-delay distance~\citep{Rusu2017}.

Meanwhile, the general form (not limited to SIS model) of the distance ratio can be expressed as~\citep{Birrer2016,Birrer2019}:
\begin{equation}
\frac{D_{ls}^A}{D_s^A}=\frac{c^2J(\boldsymbol{\xi}_{lens},\boldsymbol{\xi}_{light},\beta_{ani})}{(\sigma^P)^2},
\end{equation}
where $\sigma^P$ is the LOS projected stellar velocity dispersion of the lens galaxy.
It provides extra constraints to the cosmographic inference.
$J$ captures all the model components computed from angles measured on the sky (the imaging) and the stellar orbital anisotropy distribution.
It can be written as a function of lens model parameters $\boldsymbol{\xi}_{lens}$,
the light profile parameters $\boldsymbol{\xi}_{light}$ and the anisotropy distribution of the
stellar orbits $\beta_{ani}$. We refer to Section 4.6 of Birrer et al.(2019) for detailed modelling related with $J$.

Thus the angular diameter distance to the lens can be given by~\citep{Birrer2016,Birrer2019}:
\begin{equation}
D_l^A=\frac{1}{1+z_l}\frac{c\Delta t}{\Delta \phi(\boldsymbol{\xi}_{lens})}\frac{c^2J(\boldsymbol{\xi}_{lens},\boldsymbol{\xi}_{light},\beta_{ani})}{(\sigma^P)^2}.\label{DA}
\end{equation}
Note that a full Bayesian analysis considering covariances between quantities should be applied when dealing with the real data.
For more details of such process, we refer to Birrer et al. (2019) and Jee et al. (2015).

It is worth noting that for gravitational lensing, it is the angle measure that matters, while the intensity measure only contributes to the signal-noise-ratio (SNR).
The cosmic opacity can change the absolute intensity but not the relative intensity, thus not biasing the distance determination.
Besides, the velocity dispersion based on spectroscopic measurements are also free of the intensity.
Therefore, we point out in this paper the distances measured by gravitational lensing should be independent of cosmic opacity.

To make a forecast on the uncertainties of $D_l^A$, one needs to repeat the process in~\citep{Yildirim2019} for each system, which proved 
$D_l^A$ can be determined with $1.8\%$ precision in the best case like the lens system RXJ1131-1231.
In a general case, the $D_l^A$ may be measured with larger uncertainties. However, such simulation is beyond this work due to its complexity.
We only make several assumptions to give the expected uncertainty level for $D_l^A$ determination.
Similar to Jee et al. (2016), we assume each quantity in Eq.\ref{DA} can be measured with percent level precision currently and in the near future.
For example, the time delays measured from light curves, the spatially resolved velocity
dispersion of the lens galaxy, the LOS mass fluctuation, the Fermat potential and the parameter J determined from highly resolved imaging.
These uncertainties will result in percent level $D_l^A$ determination. We adopt a $5\%$ uncertainty of $D_l^A$ as the benchmark which is also used in
Jee et al. (2016) and Linder (2011), and as the aim of the H0LiCOW program~\citep{H0LiCOW}.
We also consider a $10\%$ uncertainty for comparison.

Current surveys, for example, the Dark Energy Survey (DES)~\citep{Treu2018} and the Hyper Suprime-Cam Survey (HSC)~\citep{More2017},
and the upcoming Large Synoptic Survey Telescope (LSST)~\citep{OM10} are bringing us new lensed quasars.
To achieve the $5\%$ precision in $D_l^A$ determination, the lensing systems should be good enough with high-quality observations.
Firstly, the time delays should be measured within percent level uncertainty.
The Time Delay Challenge (TDC) program showed
only 400 well-measured time delays are available~\citep{Liao2015b}, though LSST itself promises to find ten thousand lensed quasars~\citep{OM10}.
Furthermore, to obtain the distance information, ancillary data in terms of high-resolution imaging and spectroscopy of the lens are required.
Thus, we set the following the criteria: 1) the quasar image separation is $>1''$; 2) the third brightest image has i-band magnitude $m_i<21$;
3) the lens galaxy has $m_i<22$; 4) quadruply imaged lenses which carry more information to break the
Source-Position Transformation (SPT)~\citep{Schneider2014}.
Under these conditions, we select lenses in the OM10 catalog~\citep{OM10}\footnote{https://github.com/drphilmarshall/OM10} which is widely
used by the community in LSST lensing forecast.
55 high-quality lenses are supposed to fit these conditions. We plot the lens redshift distribution of these lenses in Fig.\ref{zl} and the uncertainty levels
in Fig.\ref{sl}.

\section{Luminosity distances from gravitational waves}
Recent detections of gravitational waves (GWs) from $\sim10$ binary black hole (BBH) mergers~\citep{GW150914,GW151226,GW170104}
and a binary neutron star (BNS) merger~\citep{GW170817} opened a new window for observing the Universe.
Especially, the electromagnetic (EM) counterparts of the BNS have been observed in wide wavelength range.
Gravitational waves from binary star mergers were proposed as the standard sirens~\citep{Schutz1986}.
The chirping GW signals from inspiralling and merging compact binary stars are self-calibrating and the luminosity
distances can be directly inferred from the detected waveforms using matched-filter method.
For illustration purposes, the strain of a chirp waveform $h\propto M^{5/3}f^{2/3}D_L^{-1}$, where $M$ is the chirp mass of the binary system,
$f$ is the frequency. $\dot{f}\propto M^{5/3}f^{11/3}$. Therefore, measuring $h, f, \dot{f}$ from the waveform can determine the luminosity distance $D_L$~\citep{Nissanke2010}.

For cosmological studies, one should also know the redshift of the GW source.
However, the GW itself does not carry the information of redshift (unless considering the tide effect~\citep{Messenger2012}).
An effective way to obtain the redshift is from the EM counterparts, for example, the
short gamma ray burst (SGRB) which is one of the most promising EM counterparts
of BNS. Once it is confirmed, the redshift can be measured from its host galaxy or afterglow.

Next generation of GW detectors, for example, the Einstein Telescope (ET)~\citep{Abernathy2011} will broaden the accessible volume
of the Universe by three orders of magnitude promising tens to hundreds of thousands of detections per year.
The detection can reach $z_{GW}=5$ with $SNR>8$. For simulating the mock data, we follow the work by Cai $\&$ Yang (2016), Zhao $\&$ Wen (2018) and Wang $\&$ Wang (2019).
The source redshift distribution follows
\begin{equation}
P(z_{GW})\propto \frac{4\pi \chi^2(z_{GW})R(z_{GW})}{H(z_{GW})(1+z_{GW})},
\end{equation}
where $\chi$ is the comoving distance and
\begin{equation}
R(z_{GW})=\begin{cases}
1+2z_{GW}, & z_{GW}\leq 1 \\
\frac{3}{4}(5-z_{GW}), & 1<z_{GW}<5 \\
0, & z_{GW}\geq 5.
\end{cases}
\label{equa:rz}
\end{equation}

\begin{figure}
 \includegraphics[width=9cm,angle=0]{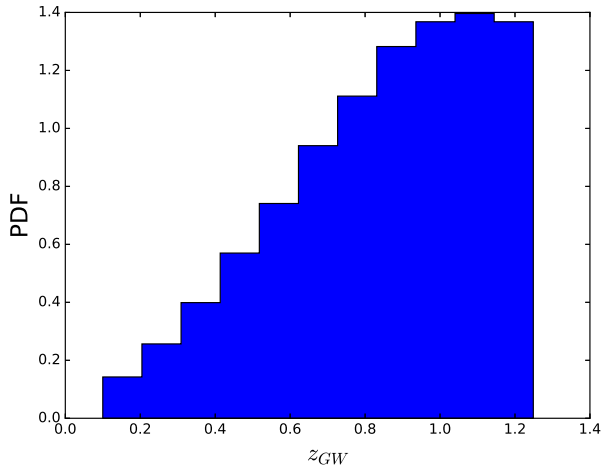}
  \caption{The redshift distribution of binary neutron stars detected by ET.
  }\label{zgw}
\end{figure}

\begin{figure}
 \includegraphics[width=9cm,angle=0]{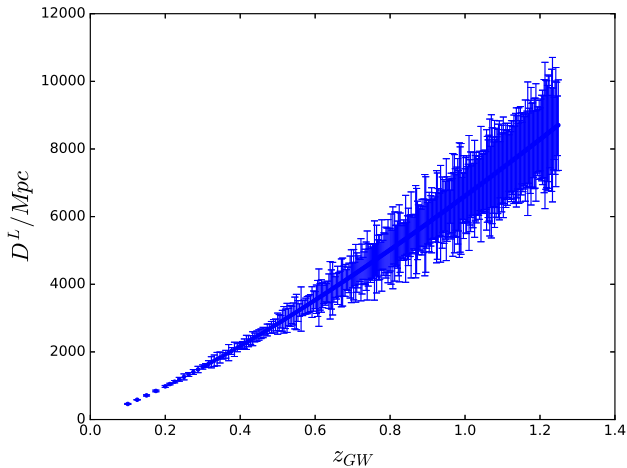}
  \caption{The simulated luminosity distances with uncertainty levels based on ET.
  }\label{gw}
\end{figure}

\begin{figure}
 \includegraphics[width=9cm,angle=0]{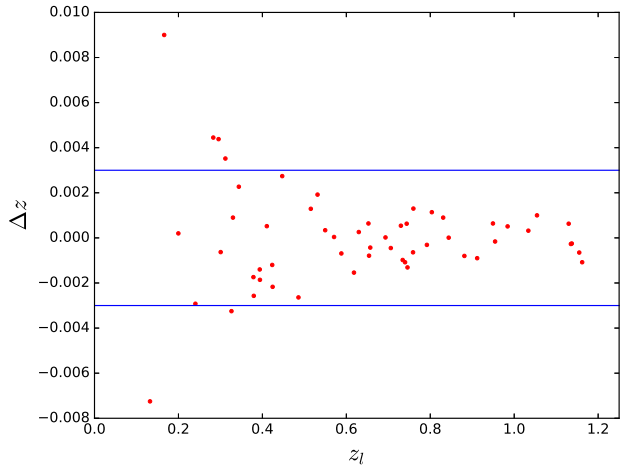}
  \caption{The redshift difference between the lens ($z_l$) and the nearest GWs ($z_{GW}=z_l+\Delta z$).
  }\label{dz}
\end{figure}

Since SGRBs are strongly beamed, only the nearly face-on configurations can provide
the redshifts, the probability is $\sim10^{-3}$, and the ET is supposed to detect $\sim10^2$
signals with accurate redshifts. For a nearly face-on system, the instrumental uncertainty
is given by
\begin{align}
\sigma_L^{\rm inst}\simeq \sqrt{\left\langle\frac{\partial \mathcal H}{\partial D^L},\frac{\partial \mathcal H}{\partial D^L}\right\rangle^{-1}},
\end{align}
where the angle bracket denotes the inner product.
$\mathcal H \propto 1/D^L$ is the Fourier transform of the waveform, then $\sigma_L^{\rm inst}\simeq D^L/\rho$, where $\rho$ is the combined SNR,
determined by the square root of the inner product of $\mathcal H$. For detecting a GW signal, $\rho>8$ is usually taken as the minimum requirement.
The uncertainty of the inclination $\iota$ would also affect the SNR, for example, the SNR would be changed by a factor of 2 from $\iota =0^{\circ}$ to $\iota = 90^{\circ}$.
Therefore, we set the instrumental uncertainty of the luminosity distance:
\begin{align}
\sigma_L^{\rm inst}\simeq \frac{2D^L}{\rho}.
\label{sigmainst}
\end{align}
In addition to the instrumental uncertainty, weak lensing effect by large-scale structure is another important systematics especially for high-redshift sources.
Ignoring them would result in biased distance measurements. Following Zhao et al.(2011),
$\sigma_L^{wl}/D^L=0.05z_{GW}$ is adopted in our simulation.
The total uncertainty on the luminosity distance is given by:
\begin{equation}
\sigma_L=\sqrt{(\sigma_L^{\rm inst})^2+(\sigma_L^{wl})^2}.
\end{equation}

To test CDDR, we choose the GWs whose redshifts $z_{GW}<1.25$ such that they can match the lens redshifts.
In Cai $\&$ Yang (2016), the total number is assumed to be 100 or 1000 up to $z_{GW}=5$, while
Zhao $\&$ Wen (2018) assumed 1000 detections within $z_{GW}=2$. Note that most of the detected sources are
at low redshifts. In this work, we take 300 sources within $z_{GW}<1.25$. The redshift distribution and the
corresponding luminosity distance uncertainty levels are shown in Fig.\ref{zgw} and Fig.\ref{gw}.

\section{methodology and results}
To test any deviation of CDDR, we parameterize it with $(1+z)^2D^A/D^L=\eta(z)$. Any $\eta(z)\neq1$ would challenge
the validation of CDDR. Since the test is only applied at low redshift $z<1.25$ and following the literature~\citep{Yang2013,Wu2015,Liao2016,Li2018,Holanda2017,Ruan2018}, we Taylor expand $\eta(z)$ in two ways:
1: with $z$, $\eta(z)=1+\eta_0z$; 2: with the scale factor $a=1/(1+z)$, $\eta(z)=1+\eta_1z/(1+z)$.

\begin{table*}\centering
 \begin{tabular}{lcc}
  \hline\hline
   Data & $\eta_0$  & $\eta_1$\\
  \hline
  (A) SNe Ia + Galaxy Clusters   &  $0.16_{-0.39}^{+0.56}$(E) or $0.02_{-0.17}^{+0.20}$(S) & $\square$\\
  (B) SNe Ia + BAOs  & $0.027\pm0.064$ & $0.039\pm0.099$\\
  (C) SNe Ia + Strong Lensing  ($D_{ls}^A/D_s^A$) & $-0.005_{-0.215}^{+0.351}$ & $\square$\\
  (D) SNe Ia + Ultra-compact Radio Sources  & $-0.06\pm0.05$ & $-0.18\pm0.16$\\
  (E) SNe Ia+GRBs+Strong Lensing ($D_{ls}^A/D_s^A$)  & $0.00\pm0.1$ & $-0.36_{-0.42}^{+0.37}$\\
  (F) HIII Galaxies + Strong Lensing ($D_{ls}^A/D_s^A$)  & $0.0147_{-0.066}^{+0.056}$ & $\square$\\
  GWs + Strong Lensing ($D^A$) (This work with $\sigma_A/D^A=5\%$) & $\pm0.026$ & $\pm0.047$\\
  GWs + Strong Lensing ($D^A$) (This work with $\sigma_A/D^A=10\%$) & $\pm0.034$ & $\pm0.058$\\
  \hline\hline
 \end{tabular}
 \caption{Comparisons with current opacity-dependent tests on CDDR. A:~\citep{Yang2013}; B:~\citep{Wu2015}
            C:~\citep{Liao2016}; D:~\citep{Li2018}; E:~\citep{Holanda2017}; F:~\citep{Ruan2018}.
}\label{result3}
\end{table*}

For a CDDR test, in principle the measured luminosity distance and angular diameter distance should correspond to the
same redshift. However, since the two distances are from different systems, their redshifts can not always be matched perfectly.
One way to deal with this is to find the nearest data pair, if the redshift difference is small enough, then they can be taken
as from the same redshift. In the literature, one usually take the $\Delta z<0.005$ as criterion~\citep{Qi2019,Liao2016,Yang2013,Li2011} and the simulations showed
this would bring ignorable systematic errors. In this work, we adopt a stricter criterion $\Delta z<0.003$. This value
is chosen such that we will still have enough matched pairs. Fig.\ref{dz} is from one of the simulations. One can see under our
criterion, there will be $\sim50$ data pairs available.

Since this work aims at giving a prediction of constraint on $\eta(z)$ rather than using realistic data to make a conclusion, we adopt two random
processes to give an unbiased result reflecting an average constraining power.
Firstly, we randomly select the redshifts of lensing and GWs; secondly, for each selected dataset,
we distribute different noise realizations to
generate the mock data. For each mock data, we do minimization using the ``scipy.optimize.minimize" function in Python to find the best-fits of $\eta_0$ and $\eta_1$. The statistic quantity used in the
minimization process is given by
\begin{equation}
\chi^2=\sum_i\frac{\left[D_i^L-D_i^A(1+z_i)^2\eta(z_i) \right]^2}{\sigma_{L,i}^2+\sigma_{A,i}^2(1+z_i)^4\eta(z_i)^2}.
\end{equation}
The total error at the denominator consists of errors from GW and strong lensing contributions. Since the typical values are both several percents across the redshift range, their contributions are comparable.

\begin{figure}
 \includegraphics[width=9cm,angle=0]{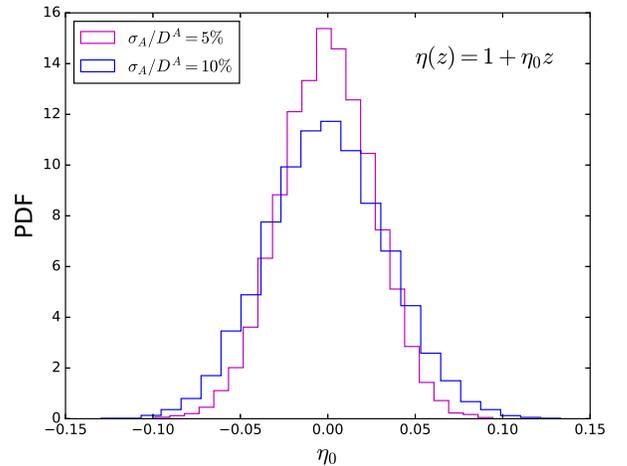}
  \caption{PDFs of the parameter $\eta_0$ assuming the uncertainties of $D_l^A$ are $5\%$ and $10\%$, respectively.
  }\label{result1}
\end{figure}

\begin{figure}
 \includegraphics[width=9cm,angle=0]{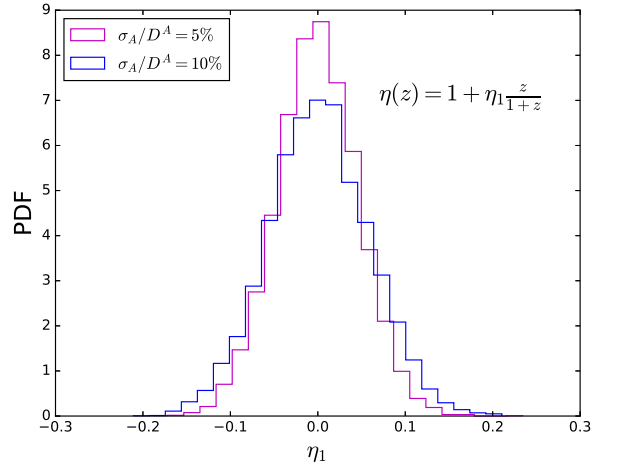}
  \caption{The same as Fig.\ref{result1} but for $\eta_1$.
  }\label{result2}
\end{figure}

\begin{figure}
 \includegraphics[width=9cm,angle=0]{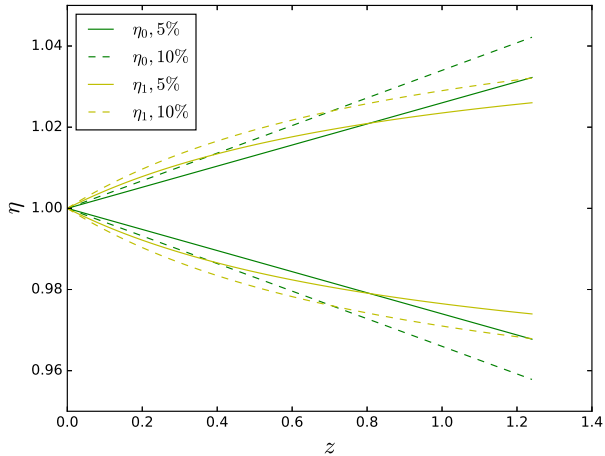}
  \caption{The reconstructed $\eta(z)$ for the two parameterizations with their errors.
  }\label{eta}
\end{figure}

We take all the best-fits from each minimizations as the expected distributions of $\eta_0$ and $\eta_1$, and plot their Probability Distribution Functions (PDFs)
in Fig.\ref{result1} and Fig.\ref{result2}, respectively. They stand for the constraining power on the deviation of CDDR.
Since the PDFs are approximately Gaussian-like,
we calculate the standard deviations as the $1\sigma$ uncertainty levels. The numerical results are summarized in Tab.\ref{result3} along
with part of the results from current methods for comparison. We also plot the reconstruction of $\eta(z)$ for the two parameterizations with their errors in Fig.\ref{eta}.
Therefore, while the CDDR test is opacity-free,
our method should be very competitive to constrain the deviation parameters. We emphasize again we do not attempt to make a constraint on CDDR from the realistic data in this work, but to propose a new method and discuss its power basing on mock data basing on future observation conditions. Therefore, only the errors matter and the best-fits are meaningless.

\section{conclusion and discussions}
We propose an opacity-free test of the cosmic distance duality relation with strong lensing and gravitational waves.
The former provides the angular diameter distance while the latter provides the luminosity distances.
Both distances are measured without the impact of cosmic opacity. This advantage ensures that we can directly test the nature of space-time by checking the validation of CDDR.

We note that the data analysis process in gravitational waves usually stands on the assumption of general relativity (GR), i.e., the GR templates.
In a modified gravity, the luminosity distance can be different~\citep{Belgacem2018}. Besides, strong lensing observation
may also be affected by the validity of GR~\citep{Collett2018}. Further works need to be done to see whether this method can
directly test GR.

During our simulations, we adopted the flat $\Lambda$CDM model with certain value of Hubble constant which is currently suffering from the tension between early-time and late-time observations~\citep{Verde2019}. However, since testing CDDR is a totally cosmology-free experiment, this tension would not affect our main results.
Assuming different values of $H_0$, for example, $67km/s/Mpc$ or $74km/s/Mpc$ would only bring in $\sim10\%$ difference for the error of $\eta$ in the estimate process.

It is worth noting that the work by Yang et al. 2019 also tested CDDR with strong lensing and GWs.
This work is different from theirs. We use the direct $D^A$ measurements from strong lensing, whereas Yang et al. 2019 used the distance ratio measurements (just a dimensionless quantity). Their idea was to extract angular diameter distance information from the distance ratios, thus the angular diameter distances are got in an indirect manner.
However, when using the distance ratio data to infer the distances, one has to assume a flat Universe in FLRW metric such that the Distance Sum Rule (DSR) can be applied $D_{ls}=D_s-(1+z_l)D_l/(1+z_s)$~\citep{Rasanen2015,Liao2016,Liao2019}. Therefore, to some degree, Yang et al. 2019 actually tested the flatness of the Universe rather than a more profound space-time nature. Also noted by us is the work by Rana et al. 2017, they used 12 realistic time delay lenses to infer the $D^A$, and compared them with $D^L$ from SNe Ia.
However, their work is cosmic-opacity-related and SNe Ia only give relative distances being suffering the calibration problems.
In this work, we use the direct $D^A$ measurements in the spirit of CDDR test (i.e., directly comparing $D^L$ with $D^A$ and testing the three conditions above) and demonstrate the $D^A$ from lensing is opacity-free, focusing on disentangling the effect of space-time nature from that of astropartical physics.

The cosmic opacity makes the observed flux changed by a factor of $e^{-\tau(z)}$, where $\tau(z)$
is the optical depth. Current mechanisms mentioned above assume $\tau(z)>0$ and increases with $z$, i.e.,
the photon number decreases rather than increases when the light passes through the Universe.
The luminosity distance will look larger than expected.
For mechanisms that violate Condition 1 and 2, we do not limit the sign of $\eta_0$ and $\eta_1$.
Among various gravity theories that violate CDDR, one may conjecture that one possibility could
be related with the non-conserved gravitons.

To give a robust test, while the observational precision is important,
the systematic errors in each observation should be carefully dealt with.
In current strong lensing techniques, the blind analysis is being conducted to control the systematics~\citep{H0LiCOW}.
More astrophysical processes are being understood, for example, the time delays caused by microlensing~\citep{Tie2018}.
For the standard sirens, the detector calibration errors, weak lensing effects and template models are being better understood.
With the developments of these two aspects, we will be able to opacity-freely test CDDR in the near future.

\section*{Acknowledgments}
The author thanks Zhengxiang Li and A. Avgoustidis for helpful comments, and Tao Yang for introducing the GW simulations.
This work was supported by the National Natural Science Foundation of China (NSFC) No. 11603015
and the Fundamental Research Funds for the Central Universities (WUT:2018IB012).

\clearpage


\begin{thebibliography}{}
\bibitem[Abbott et. al.(2016a)]{GW150914} Abbott, B. P., Abbott, R., Abbott, T. D., et al. 2016a, PhRvL, 116, 061102
\bibitem[Abbott et. al.(2016b)]{GW151226} Abbott, B. P., Abbott, R., Abbott, T. D., et al. 2016b, PhRvL, 116, 241103
\bibitem[Abbott et. al.(2017a)]{GW170104} Abbott, B. P., Abbott, R., Abbott, T. D., et al. 2017a, PhRvL, 118, 221101
\bibitem[Abbott et. al.(2017b)]{GW170817} Abbott, B. P., Abbott, R., Abbott, T. D., et al. 2017b, PhRvL, 119, 161101
\bibitem[Avgoustidis et al.(2010)]{Avgoustidis2010} Avgoustidis A., Burrage C., Redondo J., Verde L., Jimenez R. 2010, JCAP, 10. 024
\bibitem[Abernathy et al.(2011)]{Abernathy2011} Abernathy, M., Acernese, F., Ajith, P., et al. 2011, European Gravitational Observatory, Document Number ET-0106A-10
\bibitem[Bassett \& Kunz(2004)]{Bassett2004} Bassett B. A., Kunz M. 2004, PhRvD, 69, 101305
\bibitem[Birrer et al.(2016)]{Birrer2016} Birrer S., Amara A., Refregier A., 2016, JCAP, 08, 020
\bibitem[Birrer et al.(2019)]{Birrer2019} Birrer S., Treu T., Rusu C. E. 2019, MNRAS, 484, 4726
\bibitem[Bernardis et al.(2006)]{Bernardis2006} Bernardis F. D., Giusarma E., Melchiorri A. 2006, IJMPD, 15, 759
\bibitem[Belgacem et al.(2018)]{Belgacem2018} Belgacem E., Dirian Y., Foffa S., Maggiore M., 2018, PhRvD, 97, 104066
\bibitem[Chae(2003)]{Chae2003} Chae K.-H., 2003, MNRAS, 346, 746
\bibitem[Chen et al.(2019)]{Chen2019} Chen Y., Li R., Shu Y., Cao X., 2019, MNRAS accepted, arXiv: 1809.09845
\bibitem[Cao et al.(2012)]{Cao2012} Cao S., Pan Y., Biesiada M., Godlowski W., Zhu Z.-H., 2012, JCAP, 03, 016
\bibitem[Cai \& Yang(2017)]{Cai2017} Cai R.-G., Yang T. 2017, PhRvD, 95, 044024
\bibitem[Collett et al.(2018)]{Collett2018} Collett T. E., Oldham L. J., Smith R. J., et al., 2018, Science, 6395, 1342
\bibitem[Etherington(1933)]{Etherington1933} Etherington, I. M. H. 1933, Phil. Mag., 15, 761
\bibitem[Fu et al.(2019)]{Fu2019} Fu X., Zhou L., Chen J. 2019, arXiv: 1903.09913
\bibitem[Holanda et al.(2010)]{Holanda2010} Holanda R. F. L., Lima J. A., Ribeiro M. B. 2010, ApJL, 722, L233
\bibitem[Holanda et al.(2012)]{Holanda2012} Holanda R. F. L., Lima J. A., Ribeiro M. B. 2012, A\&A, 538, A131
\bibitem[Holanda et al.(2016)]{Holanda2016} Holanda R. F. L., Busti V. C., Alcaniz J. S. 2016, JCAP, 02, 054
\bibitem[Holanda et al.(2017)]{Holanda2017} Holanda R. F. L., Busti V. C., Lima F. S., Alcaniz J. S. 2017, JCAP, 09, 039
\bibitem[Hu \& Wang(2018)]{Hu2018} Hu J., Wang F.Y., 2018, MNRAS, 477, 5064
\bibitem[Jee et al.(2015)]{Jee2015} Jee I., Komatsu E., Suyu S.-H. 2015, JCAP, 11, 033
\bibitem[Jee et al.(2016)]{Jee2016} Jee I., Komatsu E., Suyu S. H., Huterer D. 2016, JCAP, 04, 031
\bibitem[Jiang \& Kochanek(2007)]{Jiang2007} Jiang G., Kochanek C. S. 2007, ApJ, 671, 1568
\bibitem[Li et al.(2011)]{Li2011} Li Z., Wu P., Yu H. 2011, ApJ, 729, L14
\bibitem[Li et al.(2013)]{Li2013} Li Z., Wu P., Yu H., Zhu Z.-H. 2013, PhRvD, 87, 103013
\bibitem[Li \& Lin(2018)]{Li2018} Li X., Lin H.-N. 2018, MNRAS, 474, 313
\bibitem[Lima et al.(2011)]{Lima2011} Lima J. A. S., Cunha J. V., Zanchin V. T. 2011, ApJ, 724, L26
\bibitem[Liao et al.(2015a)]{Liao2015a} Liao K., Avgoustidis A., Li Z. 2015a, PhRvD, 92, 123539
\bibitem[Liao et al.(2015b)]{Liao2015b} Liao K., Treu T., Marshall P., et al. 2015b, ApJ, 800, 11
\bibitem[Liao et al.(2016)]{Liao2016} Liao K., Li Z., Cao S., et al. 2016, ApJ, 822, 74
\bibitem[Liao et al.(2019)]{Liao2019} Liao K., 2019, ApJ, 833, 3
\bibitem[Linder(2011)]{Linder2011} Linder E. V., 2011, PhRvD, 84, 123529
\bibitem[Messenger \& Read(2012)]{Messenger2012} Messenger C., Read J. 2012, PhRvL, 108, 091101
\bibitem[More et al.(2017)]{More2017} More A., Lee C.-H., Ogure M., et al., 2017, 465, 2411
\bibitem[Nissanke et al.(2010)]{Nissanke2010} Nissanke S., Holz D. E., Hughes S. A., Dalal N., Sievers J. L., 2010, ApJ, 725, 496
\bibitem[Oguri(2007)]{Oguri2007} Oguri M., 2007, ApJ, 660, 1
\bibitem[Oguri et al.(2012)]{Oguri2012} Oguri M., Inada N., Strauss M. A., et al., 2012, ApJ, 143, 120
\bibitem[Oguri \& Marshall(2010)]{OM10} Oguri M., Marshall P. J. 2010, MNRAS, 405, 2579
\bibitem[Paraficz \& Hjorth(2009)]{Paraficz2009} Paraficz D., Hjorth J., 2009, A\&A, 507, L49
\bibitem[Paraficz \& Hjorth(2010)]{Paraficz2010} Paraficz D., Hjorth J., 2010, ApJ, 712, 1378
\bibitem[Qi et al.(2019)]{Qi2019} Qi J.-Z., Cao S., Zheng C., et al. 2019, PhRvD, 99, 063507
\bibitem[Rusu et al.(2017)]{Rusu2017} Rusu C. E., et al., 2017, MNRAS, 467, 4220
\bibitem[Ruan et al.(2018)]{Ruan2018} Ruan C.-Z., Melia F., Zhang T.-J. 2018, 866, 31
\bibitem[Rana et al.(2017)]{Rana2017} Rana A., Jain D., Mahajan S., Mukherjee A., Holanda R. F. L. 2017, JCAP, 07, 010
\bibitem[R\"{a}s\"{a}nen et. al.(2015)]{Rasanen2015} R\"{a}s\"{a}nen, S., Bolejko, K., Finoguenov, A. 2015, PhRvL, 115, 101301
\bibitem[Schutz(1986)]{Schutz1986} Schutz, B. F. 1986, Nature, 323, 310
\bibitem[Suyu et. al.(2017)]{H0LiCOW} Suyu, S. H., Bonvin, V., Courbin, F., et al. 2017, MNRAS, 468, 2590
\bibitem[Schneider \& Sluse(2014)]{Schneider2014} Schneider P., Sluse D. 2014, A\&A, 564, A103
\bibitem[Treu(2010)]{Treu2010} Treu T. 2010, Annu. Rev. Astron. Astrophys., 48, 87
\bibitem[Treu et al.(2018)]{Treu2018} Treu T., Agnello A., Baumer M. A., et al., 2018, MNRAS, 481, 1041
\bibitem[Tie \& Kochanek(2018)]{Tie2018} Tie, S. S., Kochanek, C. S. 2018, MNRAS, 473, 80
\bibitem[Tu et al.(2019)]{Tu2019} Tu Z. L., Hu J., Wang F. Y., 2019, MNRAS, 484, 4337
\bibitem[Uzan et al.(2004)]{Uzan2004} Uzan J.-P., Aghanim N., Mellier Y. 2004, PhRvD, 70, 083533
\bibitem[Verde et al.(2019)]{Verde2019} Verde L., Treu T., Riess A. G., 2019, arXiv: 1907.10625
\bibitem[Wu et al.(2015)]{Wu2015} Wu P., Li Z., Liu X., Yu H. 2015, PhRvD, 92, 023520
\bibitem[Wong et al.(2019)]{Wong2019} Wong K. C., Suyu S. H., Chen G. G.-F., et al., 2019, arXiv: 1907.04869
\bibitem[Wang \& Wang(2019)]{Wang2019} Wang Y. Y., Wang F. Y., 2019, ApJ, 873, 39
\bibitem[Xia et al.(2017)]{Xia2017} Xia J.-Q., Yu H., Wang G.-J., et al. 2017, ApJ, 834, 75
\bibitem[Yang et al.(2013)]{Yang2013} Yang X., Yu H.-R., Zhang Z.-S., Zhang T.-J. 2013, ApJL, 777, L24
\bibitem[Yang et al.(2019)]{Yang2019} Yang T., Holanda R. F. L., Bin H. 2019, Astroparticle Physics, 108, 57
\bibitem[Y{\i}ld{\i}r{\i}m et al.(2019)]{Yildirim2019} Y{\i}ld{\i}r{\i}m A., Suyu S H., Halkola A. 2019, arXiv:1904.07237
\bibitem[Zhao et al.(2011)]{Zhao2011} Zhao W., Van Den Broeck C., Baskaran D., Li T. G. F., 2011, PhRvD, 83, 023005
\bibitem[Zhao \& Wen(2018)]{Zhao2018} Zhao W., Wen L. 2018, PhRvD, 97, 064031





\end{thebibliography}
\end{document}